\documentstyle[psfig]{elsart}
\begin{document}

\begin{frontmatter}
\title{
Experimental Information on the Mass Composition of
Primary Cosmic Rays Using the KASCADE array}

\author{Tadeusz Wibig}

\address{Experimental Physics Dept., University of \L odz,
Pomorska 149/153, 90-236 \L odz, Poland}


\maketitle

\begin{abstract}

A procedure is proposed to extract experimental
information about the primary cosmic Ray composition
from the data taken with the \protect{KASCADE} setup.
\end{abstract}

\end{frontmatter}

\section {Introduction}

The data collected with the KASCADE detector setup (Ref.~\cite{kascade})carry
information first about the electromagnetic component of
extensive air showers (EAS), sufficiently acurateto enable the reconstruction
of the shower core
position and the angle of incidence, and to determine the electron--photon
density distribution (at last at larger distances from the center) for
individual shower. For
the analysis proposed in this note only the knowledge of the electron--photon
density at a particular fixed distance $r$
will be needed. The accuracy of its determination, dependent on shower size,
core location and inclination of the shower, can be expected to be in the order
of few percent.

For the muon component data will be measured for different energy thresholds
of various moun detector devices of KASCADE.
The array registrates muons with energies aabove 100 MeV, while the
muon multiwire proportional chamber installation below the iron--sampling
calorimeter of the central detector has a muon detection threshold of 2 GeV.
In addition there is a moun tracking hodoscope setup in a tunel under
installation with the threshold of 900 MeV.
The different data sets may be analyzed separately,
providing cross checks, or in a combined way, increasing
the accuracy of the estinate muon distribution parameters. That accuracy
is certainly worse than for the electron--photon component, first of all
due to the lower densities of the muon component.
Nevertheless the uncertainty of the muon density at some intermediate distance
from the core may be not larger than about 20\%.
In addition to the muon density at the fixed distance $r$, an additional
parameter of the muon lateral
distribution can be extracted and used for the analysis.
Most promising appears the slope
of that distribution ( defined by
$-\: \left( {d \ln{\rho_\mu} \over {d \ln{r}} }\right)$
at the fixed radial distance $r$ ). The accuracy of its
determination is not a crucial for the proposed procedure of the data
analysis.

\section {Method}

For the CR showers registered by the experiment the density correlation can be
displayed in two-dimensional histogram in
$\left( \rho _{\mu}(r_0), \rho_e (r_0) \right)$, similar to the well-known
$(N_\mu, N_e)$ plot. From the experimental point of view the densities at
a fixed distance can be much better defined so the
experimental uncertainties concerning the $N_\mu$ and $N_e$ determination
vanish.

\centerline{\mbox{
\psfig{file=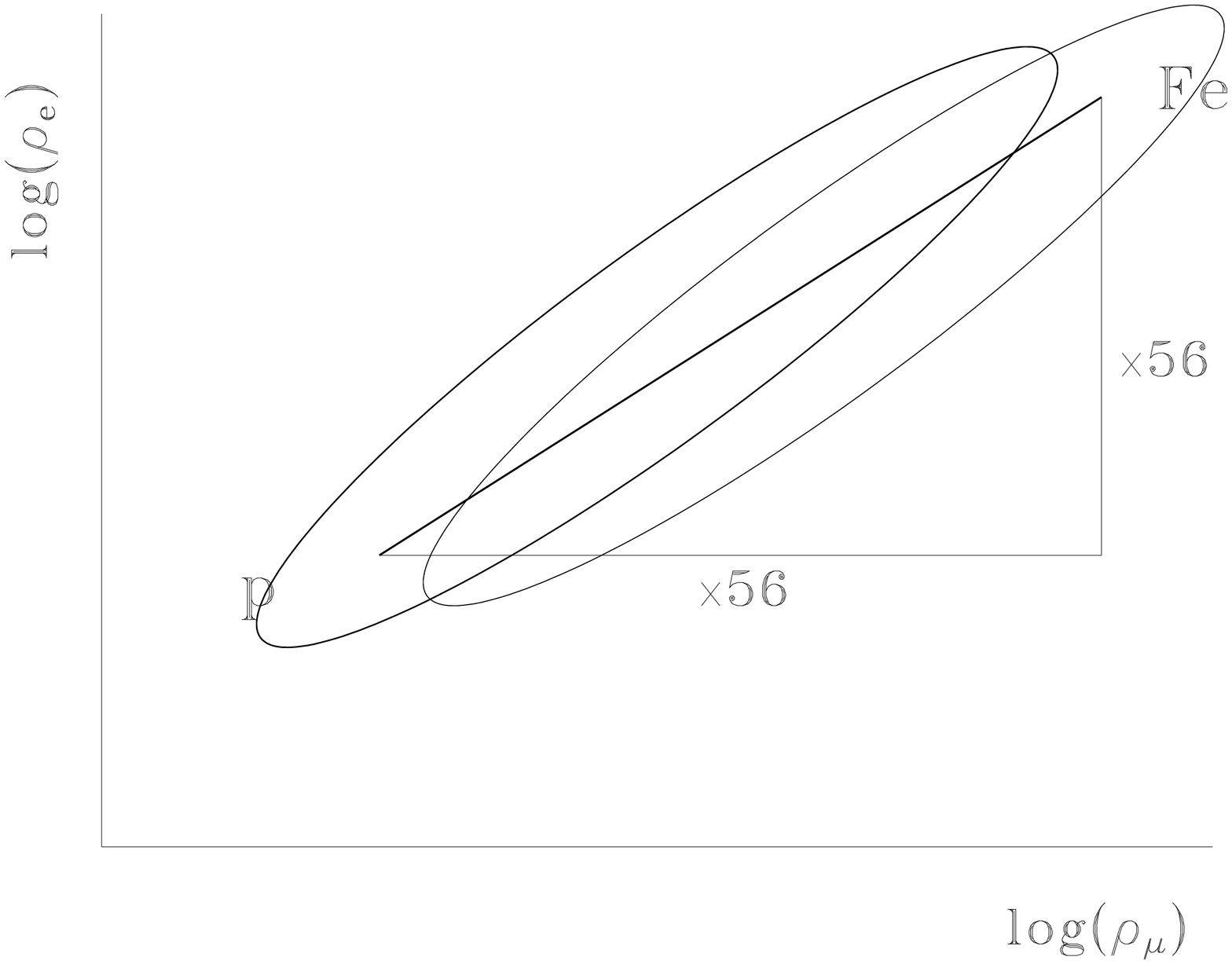,width=8cm,angle=270}}}

Fig. 1. Positions of the points representing individual showers initiated
by the primary
protons and iron nuclei in the
two-dimensional plot $\ln \left ( {\rho _{\mu}(r_0)} \right ) \times
\ln \left ( {\rho_e (r_0)} \right)$.

\vspace{1cm}

In the Fig.1 there are schematically
presented points appearing in the $\left( \rho _{\mu}(r_0), \rho_e (r_0) \right)$
plot for proton and iron induced showers. The vector
$(56 \times , 56 \times )$ gives the connection
between the two groups and it is a consequence of the compositeness of
the iron nucleus and it is very hard to change that factor much using any
sophisticated interaction model. The simple superposition model leads to the
exact $56$ value. The more realistic calculations shows that the mean values
of shower characteristics are in surprising good agreement with the
superposition assumption (Ref.~\cite{sup}) only fluctuations are larger then expected
($ 1 \ / \ \sqrt{A} \sigma_{p} $). This is quite obvious because
subshowers in the composite nucleus cascade are not independent.
Divergences in the mean values are mainly due to the effects of differences
in interaction cross-sections used and the details in nucleus-nucleus
interaction treatment.

If for each $\left( \rho _{\mu}(r_0), \rho_e (r_0) \right)$ bin the average
value of the $\mu$ distribution slope parameter will be given that for the
proton showers the clear dependence is expected.

\centerline{\mbox{
\psfig{file=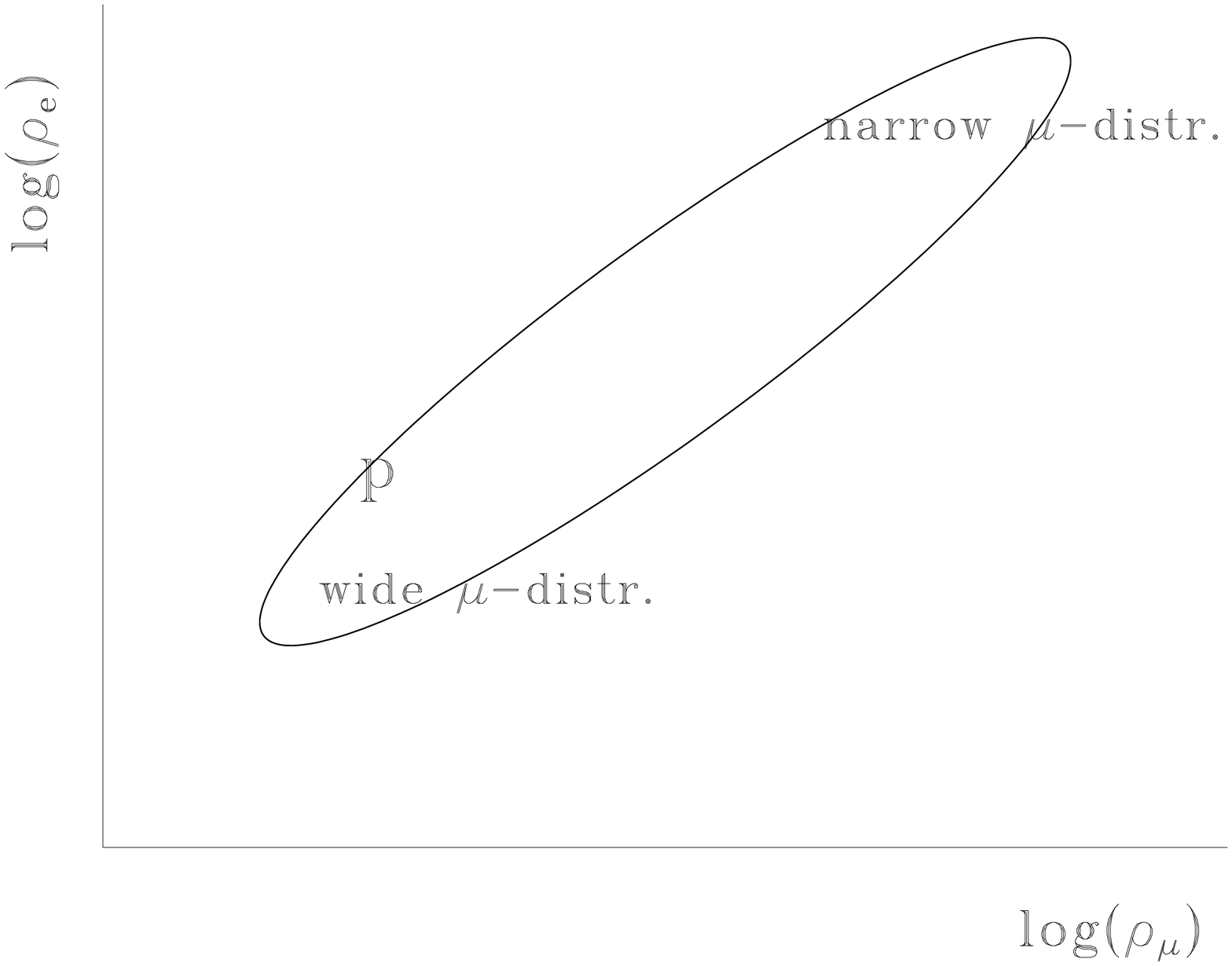,width=8cm,angle=270}}}

Fig. 2. The expected change in the $\mu$ distribution slope parameter
for the proton
induced showers of different electron and muon sizes.

\vspace{1 cm}

For the smaller showers muons has to be distributed wider than for
the larger ones.

That situation is presented in details in the Fig.8. The same dependence
is of course expected for any kind of primary particle (Fig. 9) because it
is a simple result of the geometry of the shower development. It is again
rather hard to think about the shower model without such a behaviour.
For the more energetic primary particle extensive air shower has to
developed deeper in the atmosphere so there must be an increase of the
number of muons observed on the ground generated closer to the observation
level. If no unexpected change of the interaction picture is assumed than
all that muons has to be distributed closer to the shower axis simply because
of that they have no time to wander broader.

In the Fig. 3 the situation for proton and iron initiated showers is shown.

\centerline{\mbox{
\psfig{file=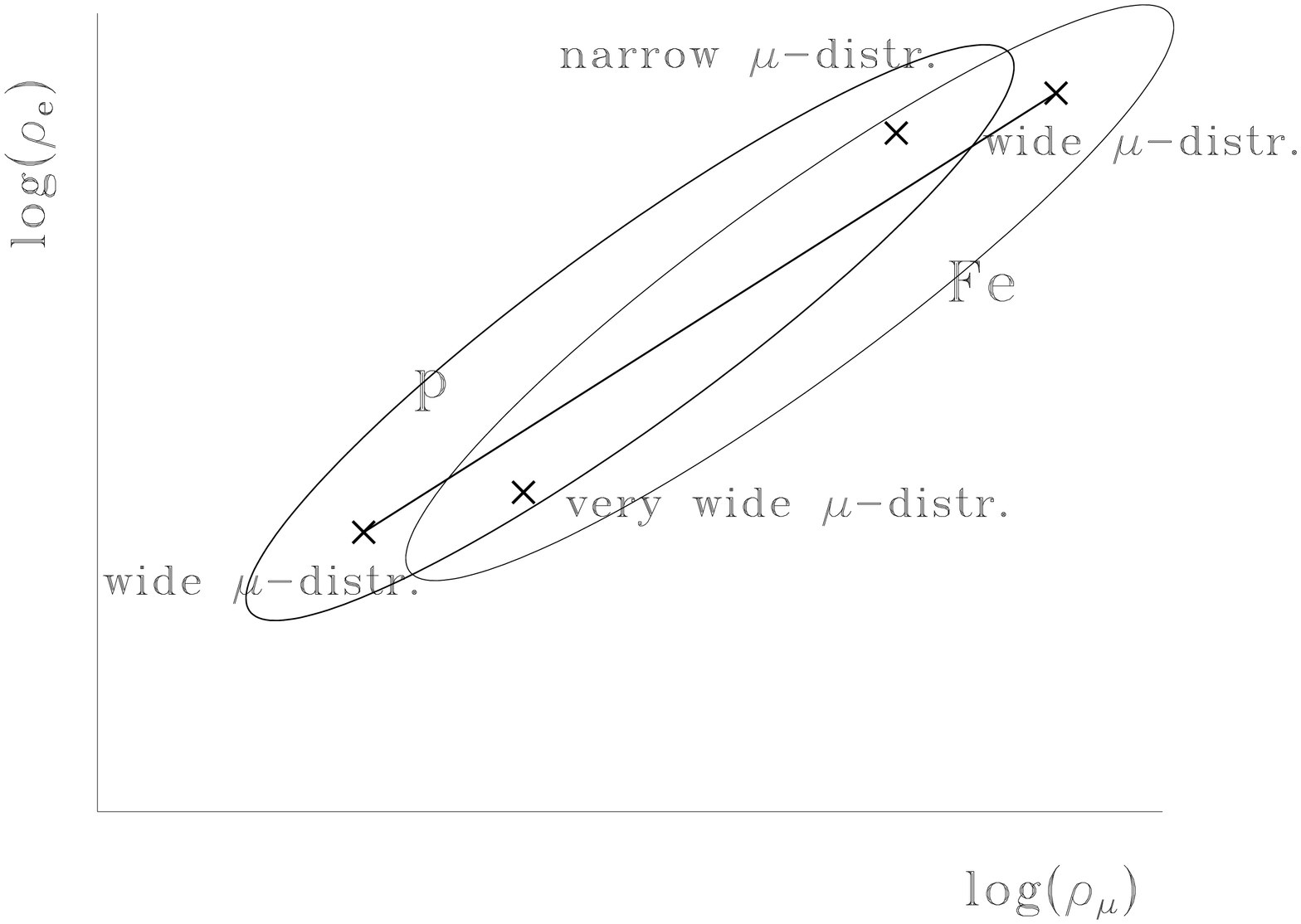,width=8cm,angle=270}}}

Fig. 3. The expected change in the $\mu$ distribution slope parameter
for the
proton and iron induced showers.

\vspace{1 cm}

If we shift the ''wide proton shower points'' by the
vector $(56 \times , 56 \times )$ we obtain the points for iron showers
with approximately the same energy per nucleon. The slope of the
$\mu$ lateral distribution has to be approximately the
same. The word ''approximately'' in the last sentence has the different
meaning that the same word in the previous one. While the first is, as
discussed above, related to the applicability of the superposition
model for the mean value of the electron and muon densities in the shover,
the second is much sensitive to the shower development fluctuations.

However the calculation shows (see Figs. 8--13, 15 and 16) that the effect
is so big that the spread due to the fluctuations of the proton shower
development for the fixed primary energy is not able to diminish it.
This is the main point of the analysis proposed.

For the primary spectrum with the one component only (e.g. protons) the slope
for fixed value of $\rho _{\mu} (r_0)$ (or $\rho _e $) should not change
very much but if we assume that there is an important
fraction of other (heavier) nuclei in the primary CR mass spectrum the $\mu$
slope has to change from the one given in Fig. 4.:

\centerline{\mbox{
\psfig{file=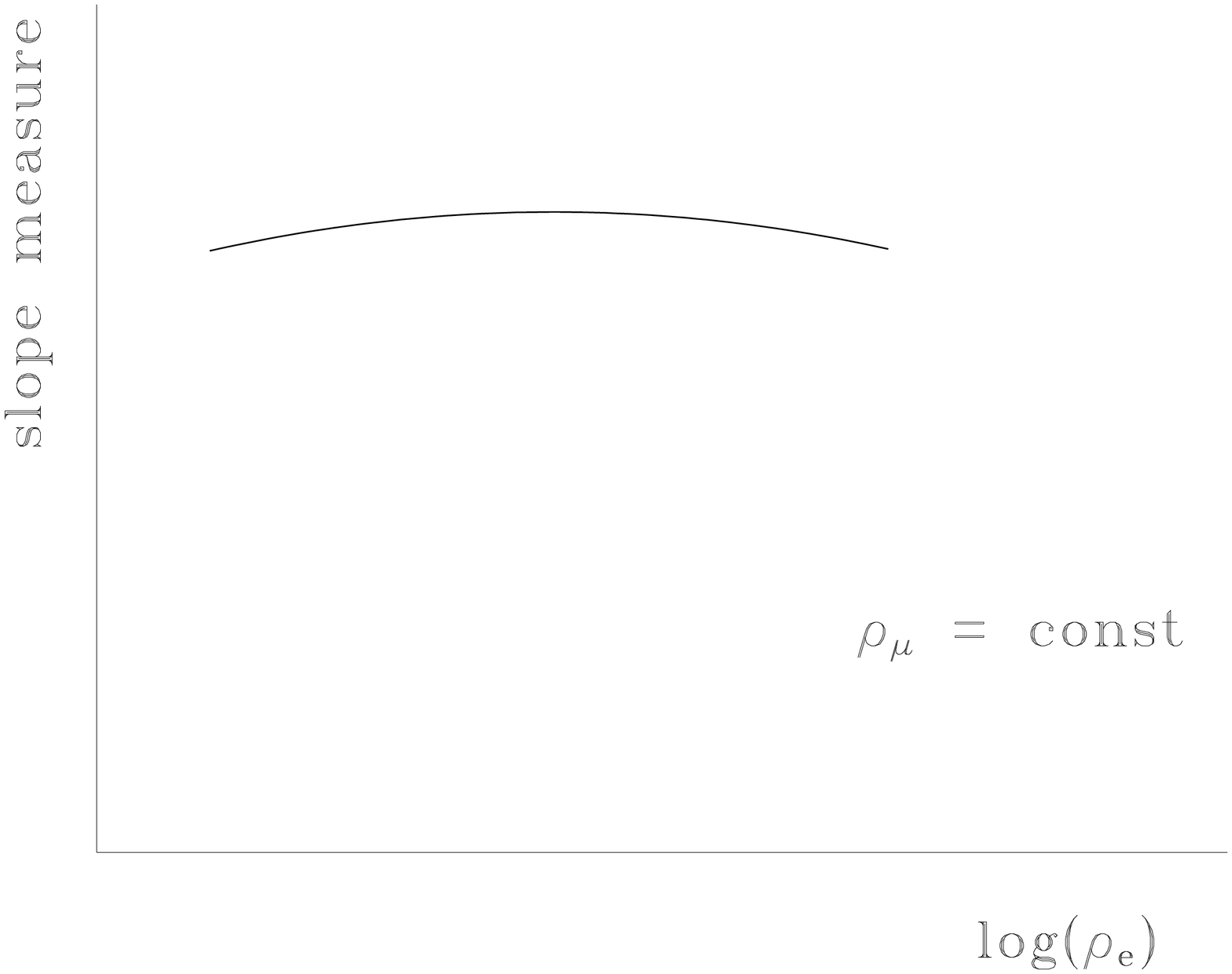,width=6cm,angle=270}}}

Fig. 4. The slope of the $\mu$ distribution measure of the proton
induced showers
as a function
of the muon density at the fixed distance from the shower
core for the
showers with the given electron density at some fixed distance.

\vspace{1 cm}

to the one presented below in Fig. 5:

\centerline{\mbox{
\psfig{file=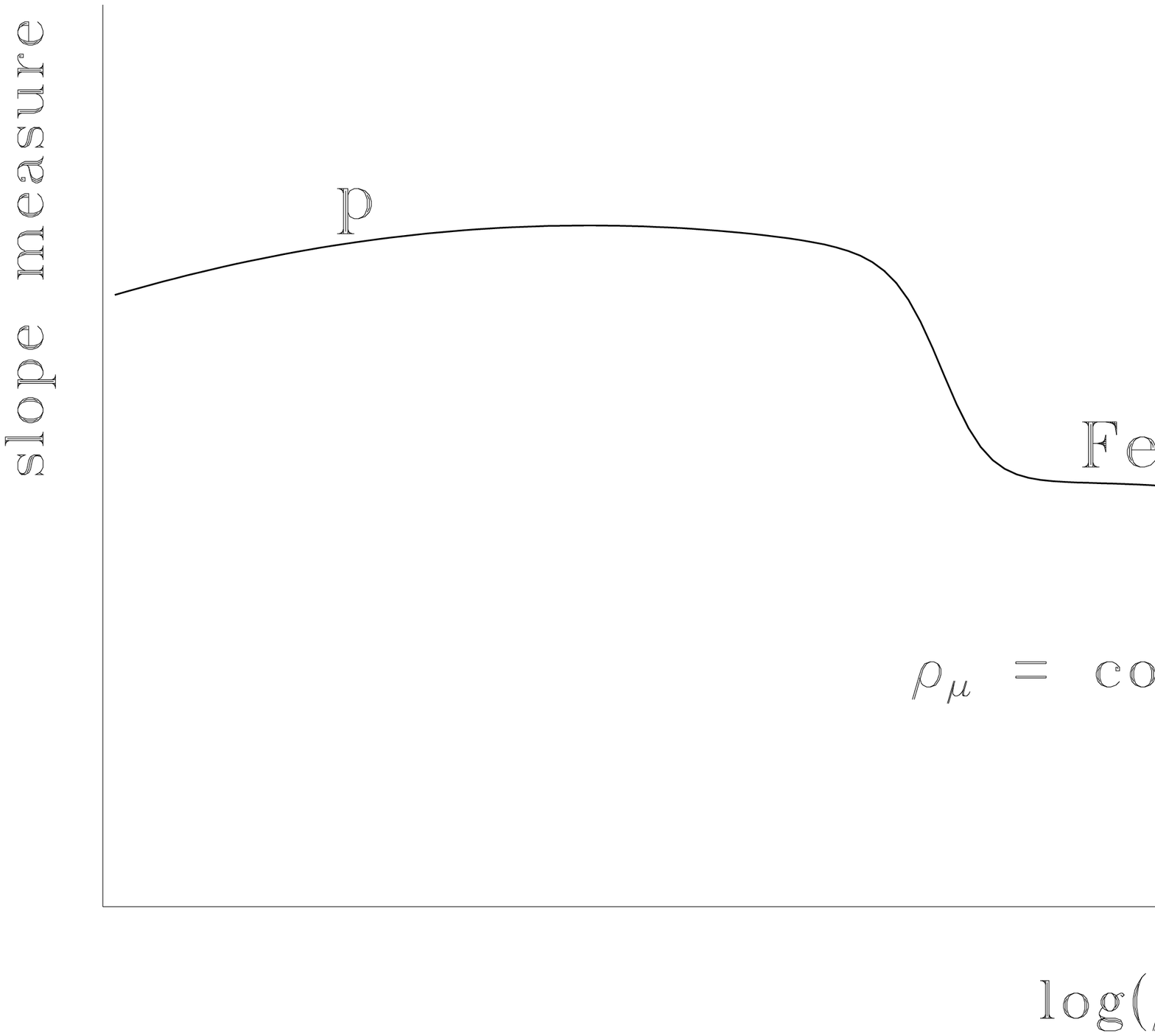,width=6cm,angle=0}}}


Fig. 5. The slope of the $\mu$ distribution measure of the proton and
iron in the
primary CR spectrum as a function
of the muon density at the fixed distance from the
shower core for the
showers with the given electron density at some fixed distance.

\vspace{1 cm}

Details can be seen in Figs. 10 and 11.

In the real case the primary CR mass spectrum is much complex
than protons and iron nuclei only. In that case there are some points on
$\left( \rho _{\mu}(r_0), \rho_e (r_0) \right)$ plot shifted from the
pure proton picture also by the vectors $(28 \times , 28 \times )$,
$(14 \times , 14 \times )$ and $(4 \times , 4 \times )$
for Ne-S, C-O and He nuclei group respectively.
This is shown schematically in Fig. 6.

\centerline{\mbox{
\psfig{file=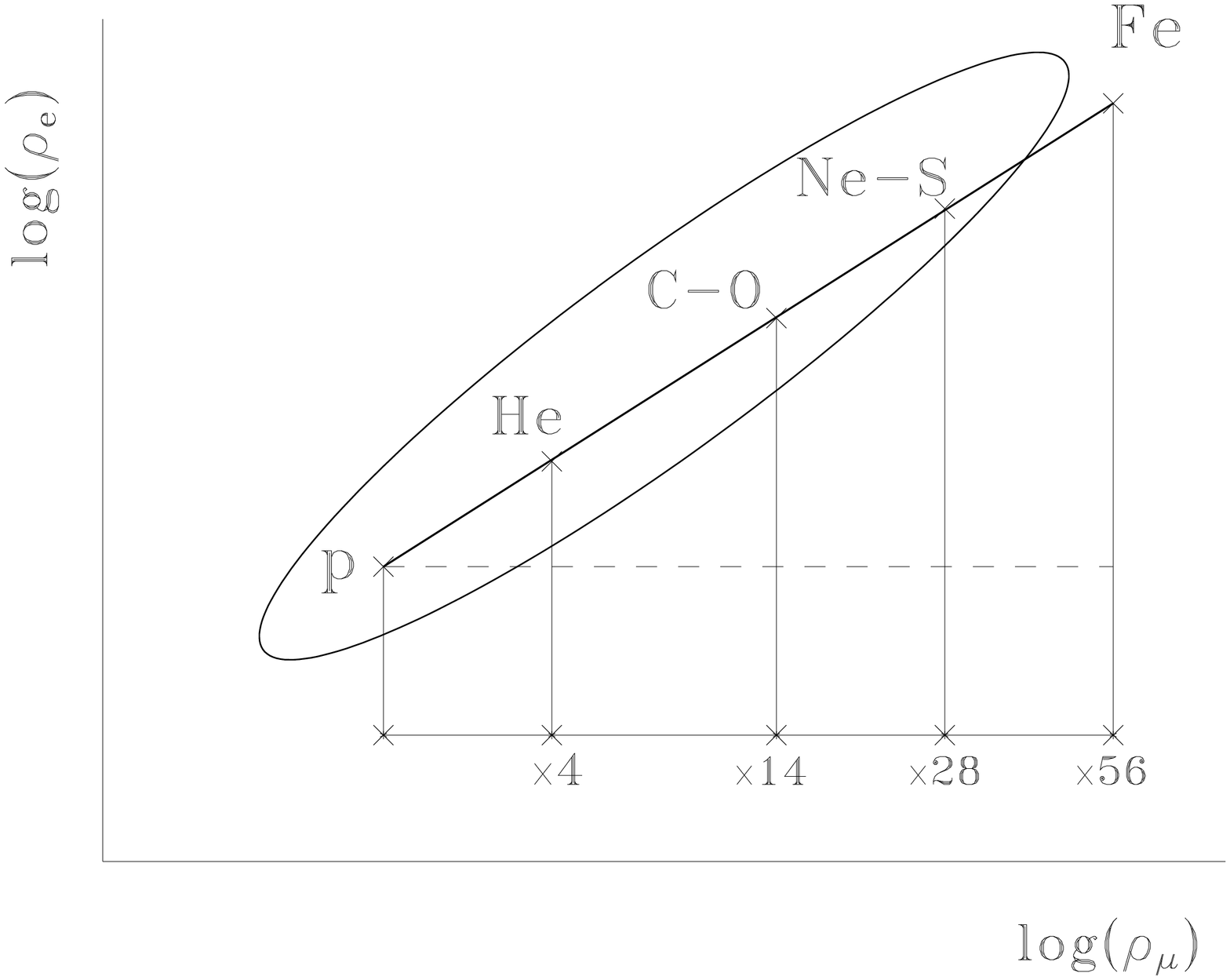,width=8cm,angle=270}}}

Fig. 6. Positions of the points representing individual showers initiated
by the primary
protons, He, C-O, Ne-S and Fe nuclei in the
two-dimensional plot $\ln \left ( {\rho _{\mu}(r_0)} \right ) \times
\ln \left ( {\rho_e (r_0)} \right)$.

\vspace{1 cm}

The ''realistic'' view is given below.

\centerline{\mbox{
\psfig{file=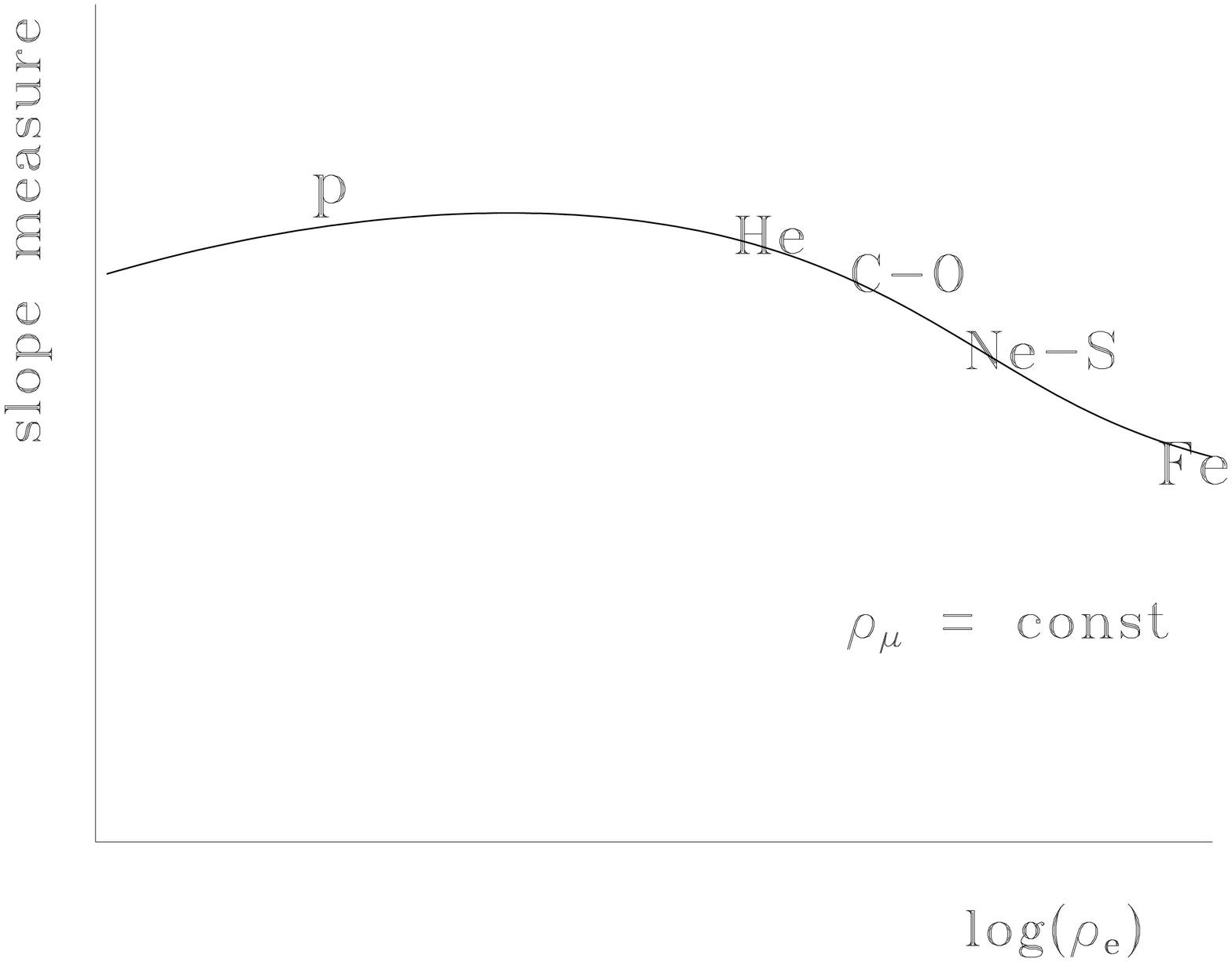,width=6cm,angle=270}}}

Fig. 7. The slope of the $\mu$ distribution measure of the proton and
iron in the primary
CR spectrum as a function
of the muon density at the fixed distance from the shower
core for the
showers with the given electron density at some fixed distance.

\vspace{1 cm}

It makes the situation from Fig. 5 less clear, anyhow points to the left
in the Fig. 6 have to give the $\mu$-slope of the
proton showers (if there are protons at all in the  primary CR spectrum)
while points the most to the right will give the $\mu$-slope of the iron
(again: if there is any).
Only the detail shape of change from the left-side to the right-side value
is related to the different nuclei abundances in primary CR.

Results of calculations are presented in Fig. 12. The mass spectrum used
was the one measured
recently by JACEE group (Ref.~\cite{jacee}) at about
$10^{14}$ eV per nucleus
$H:He:C-O:Ne-S:Fe \sim 24\%:33\%:16\%:14\%:13\%$.

The statistics of showers shown is approximately the same as for about one
week of the KASCADE data collecting. The measure of the $\mu$-slope
used is $\left( - \: {{d ln \rho_\mu} \over {d ln r}} \right )$ at
the distance from the shower core $r_0 = 50 m$.

In the Fig. 8 the results on the slope measure for pure proton spectrum is
given.
The comparison of this results with the one in Fig. 9 for pure iron
spectrum shows that the slope measure differs significantly (as it was
expected). The Figs. 10 and 11 were obtained for two different spectra
compositions consisting different amount of protons and iron nuclei in the
primary cosmic ray flux. Again as it was expected, the difference of the
muon distribution slopes seen is not a big one and can be seen only in those
$\left( \rho _{\mu}, \rho_e \right)$ bins which lay close to the
center of the area occupied by the showers were both: showers initiated by the
protons and by the iron give the contribution. On the edges the slopes
are determined by the proton (upper edge) and iron nuclei (lower edge).

The realistic situation for more complete mass composition is given in
Fig. 12.

All the Figs. 8 -- 12 were obtained using the shower generator
based on the CORSIKA simulation program (Ref.~\cite{corsika}).

\centerline{\mbox{
\psfig{file=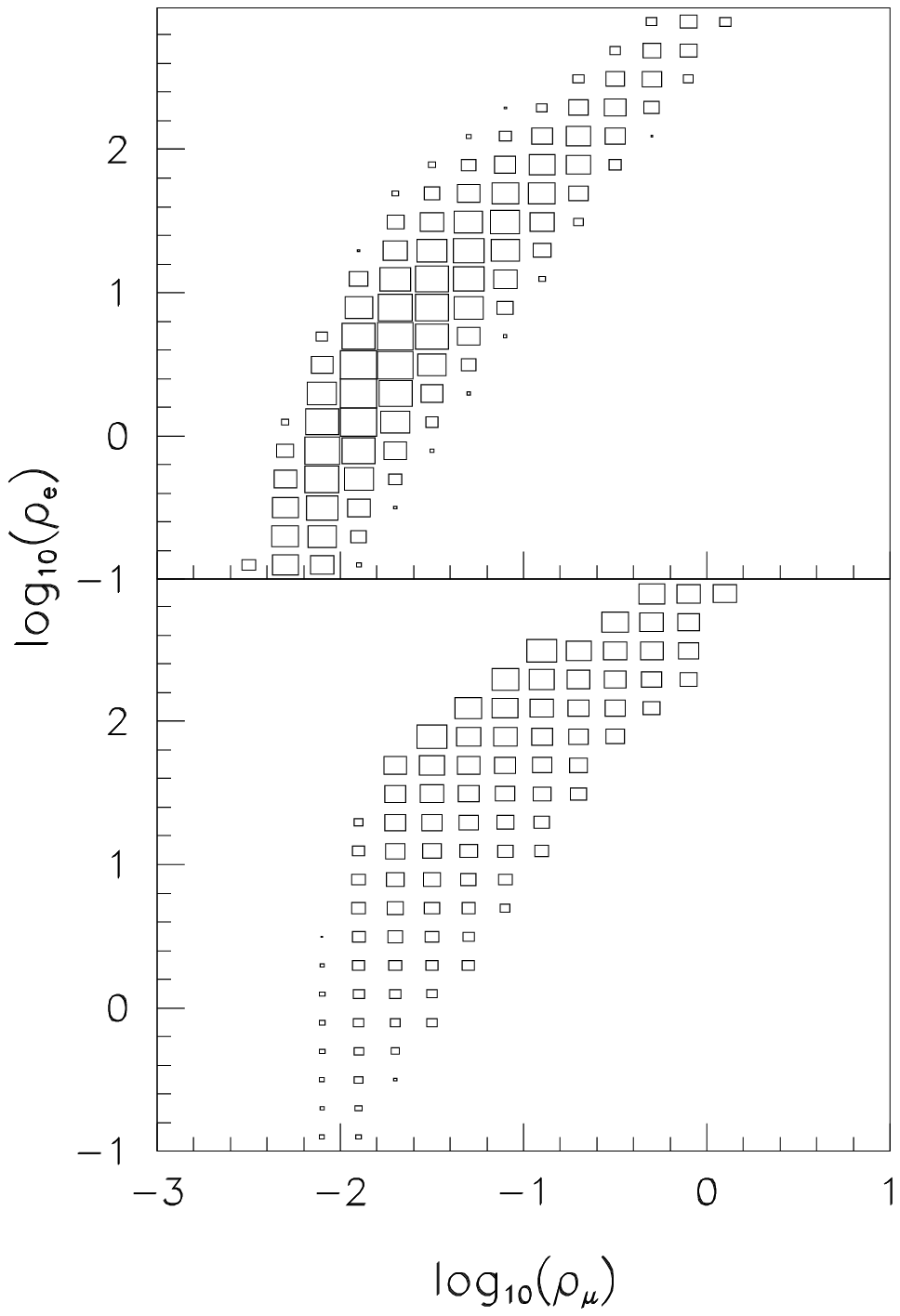,width=12cm,angle=0}
}}

Fig. 8. The results on simulation of pure proton CR primary mass spectrum.
The upper figure shows the two dimensional spectrum of
$\left( \rho _{\mu}(10m), \rho_e (50m) \right)$.
The size of each boxes is
proportional to the logarithm of the number of showers with the given electron
density at radial distance 10m and the muon density at 50m.
The lower figure presents the slope of the muon distribution at radial
distance 50m for each given $\left( \rho _{\mu}(10m), \rho_e (50m) \right)$
shower subsample. The size of each box is proportional to
$-\: \left( {d \ln{\rho_\mu} \over {d \ln{r}} }\right)$.

\centerline{\mbox{
\psfig{file=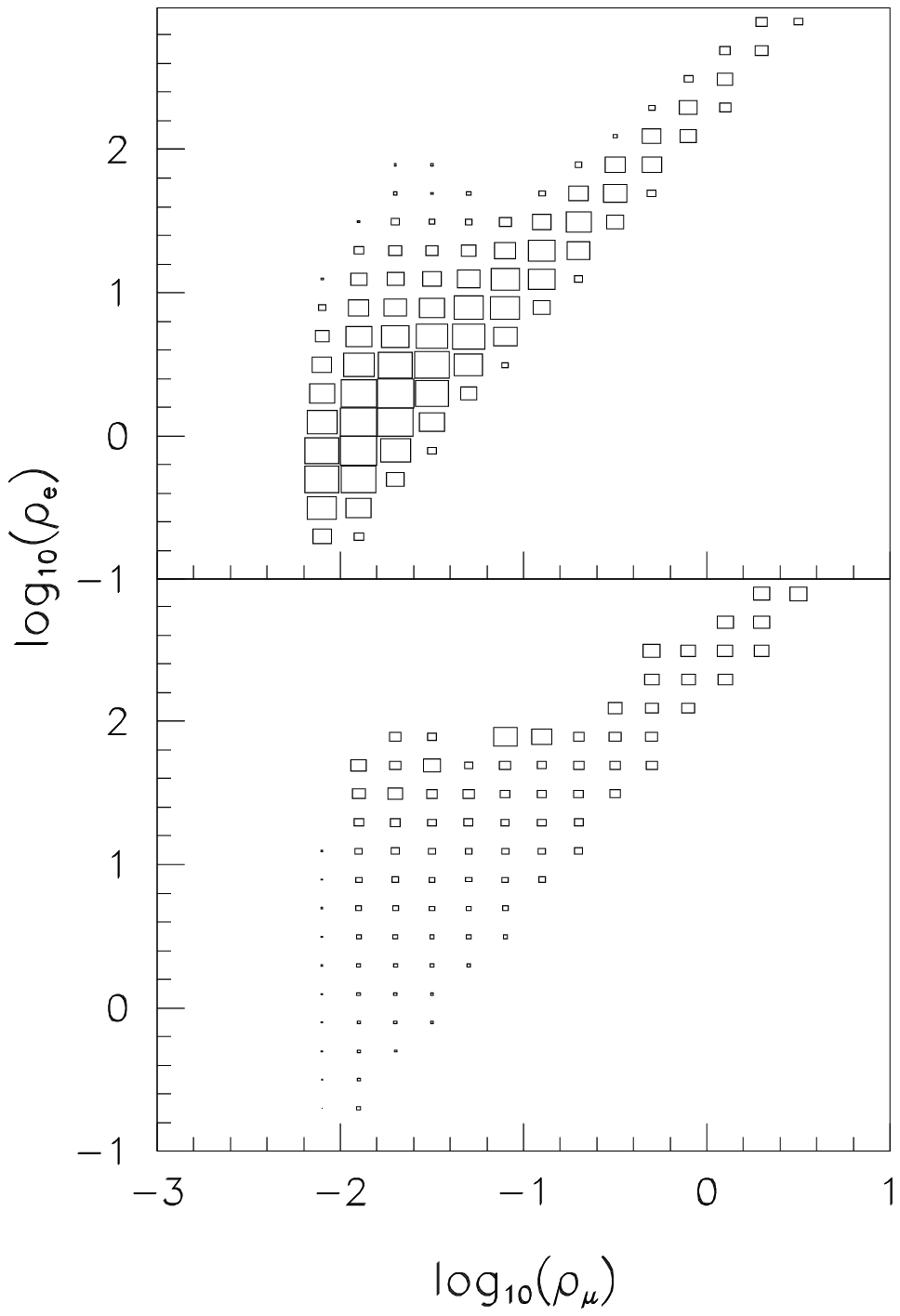,width=12cm,angle=0}
}}

Fig. 9. The results on simulation of pure iron CR primary mass spectrum.
Description as for Fig. 8.

\centerline{\mbox{
\psfig{file=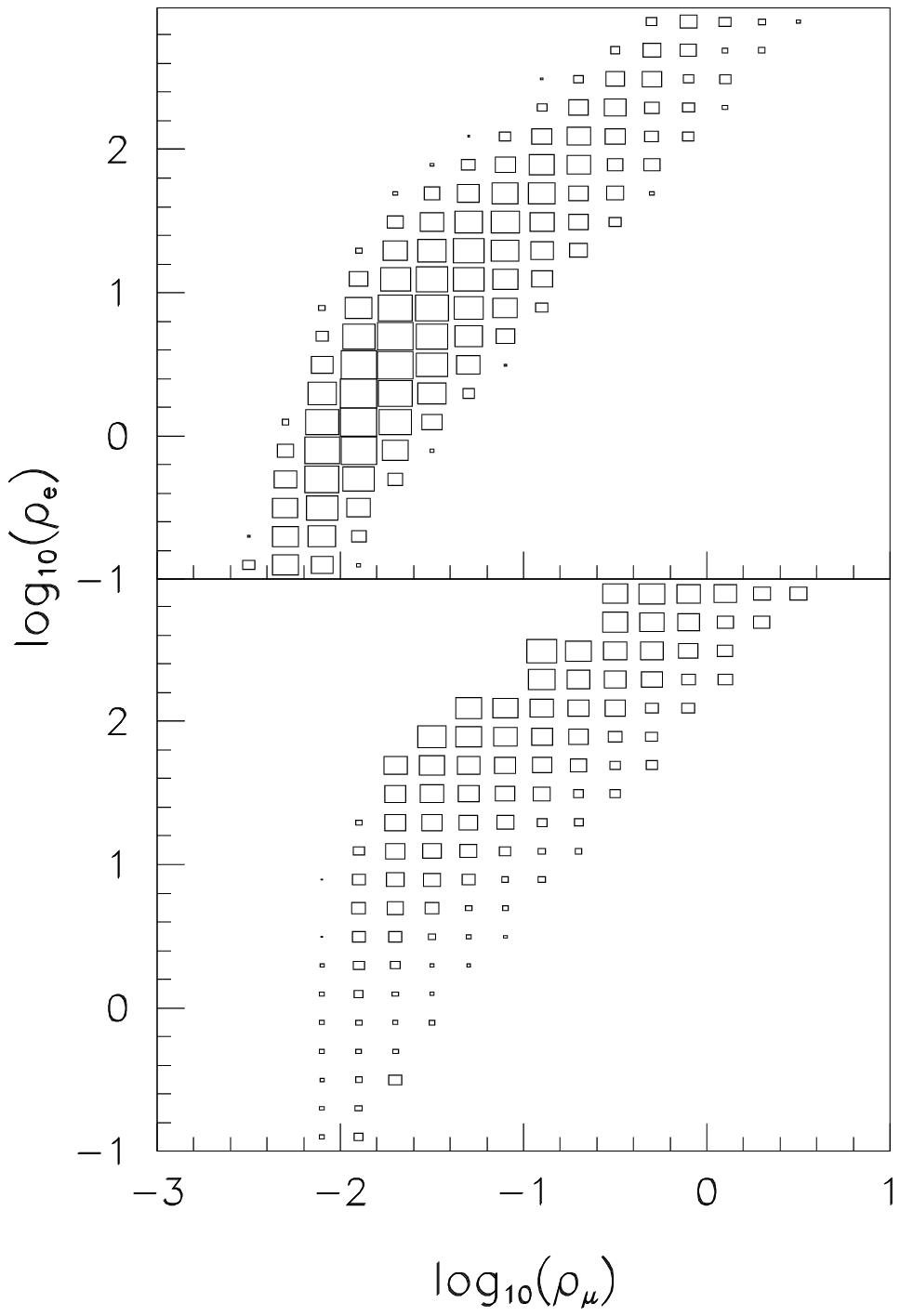,width=12cm,angle=0}
}}

Fig. 10. The results on simulation of 80\% iron and 20\% of protons
in the CR primary mass spectrum.
Description as for Fig. 8.

\centerline{\mbox{
\psfig{file=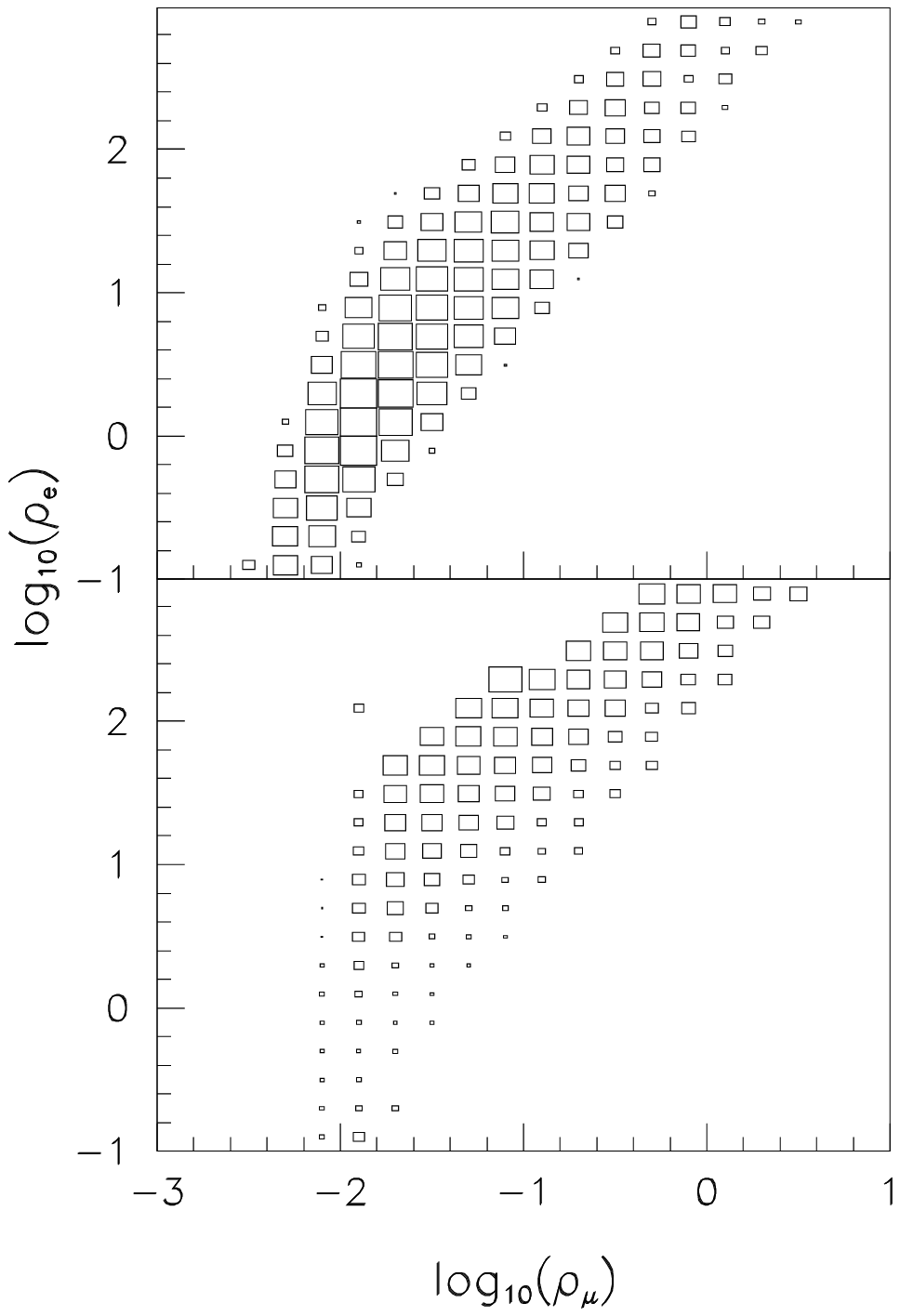,width=12cm,angle=0}}}

{Fig. 11. The results on simulation of 60\% iron and 40\% of protons
in the CR primary mass spectrum.
Description as for Fig. 8.}

\centerline{\mbox{
\psfig{file=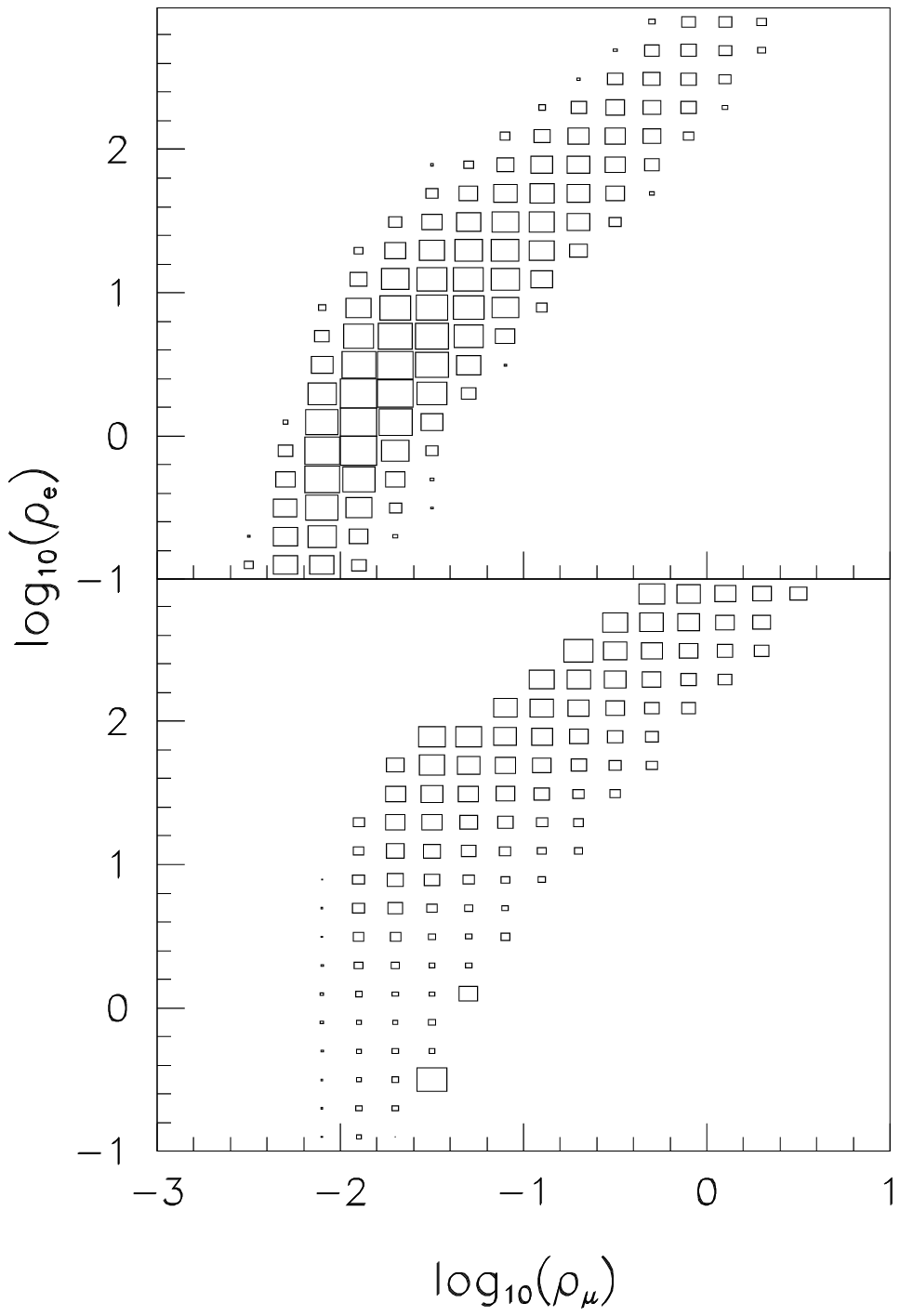,width=12cm,angle=0}}}

{Fig. 12. The results on simulation of JACEE CR primary mass spectrum.
Description as for Fig. 8.}

 \section {Summary}

Fig. 12 shows that the few weeks of the operation of the
the KASCADE experiment will be enough to proof the ability of the CR
mass spectrum determination with the method of the individual Extensive Air
Shower study.

The proposed method of the
data analysis is free of any Monte-Carlo simulation based predictions and
arguments. (It can be called ''model independent''.) Of course if one wants to
obtain a detail information about the primary mass composition ( in
principle from the plots like that in the Fig.7 it is possible ) he
ought to assume at least the $\mu$-slope as a function of $\rho_\mu$
and $\rho_e$ for proton showers and
the $\left( \rho _{\mu}(r_0), \rho_e (r_0) \right)$ proton shower spectrum
(as that like in upper part of the Fig.8) but it is ''simulation dependent'',
to some extend of course.

The analysis also will give the proof of the usefulness of the $\mu$-slope
parameter for the multiparameter individual shower study of the primary CR
mass composition in KASCADE experiment.

\section{Appendix: Accuracy of shower parameters determination}

The results presented above in Figs. 8 -- 12 were obtained only using the
shower generation algorithm without taking into account the real apparatus
performance. The important question is if the uncertainties due to the
shower reconstruction procedure for a given particular EAS array do not
introduce too much distortion to the data interpretation.
That question can be answered only by testing the data evaluation procedures.
For the KASCADE array geometry the quality of the
data collected by the large amount of detectors distributed over the
${\rm 200m \times 200m}$ area is expected to be good enough for a discussed
above purposes specially for the bigger showers.
In the Fig. 13 there are given the accuracies of shower
parameter determination important for the purposes discussed above.

It can be seen that for small showers (primary energy of about $10^{16} eV$)
the accuracy of the muon slope parameter defined as

\begin{equation}
\alpha (r_0) =  -
{\left . {{d ln \rho_{\mu}} \over {d ln r}} \right |}_{r=r_0}
\end{equation}

is rather poor. But the
$10^{17} eV$ showers are big enough to be used directly as it was proposed.
However for the small showers the situation is not hopeless. The muon
distributions for each $(\rho_{e},\rho_{\mu})$ bin can be
averaged for many showers and after that the slope determination can
be performed.

In the Figs. 14 and 15 there are 3-dim plots like that in Figs. 8 -- 12
for the showers used for the quality test in the Fig. 13.
Upper figures shows again the number of showers in each
$(\rho_{e}(50),\rho_{\mu}(50))$ bin (the size of the box in proportiona l to
the logarithm of the number of hits) and the lower shows the average
slope in each bin (here the size of the box is proportional to the
$\alpha (50m)$). The plots in the Fig. 14 shows the
{\it true} shower values while the results obtained by the evaluation
of the detected particle numbers {\it reconstructed events}
are given in the Fig. 15. As it can be seen for
the $10^{17} eV$ showers the
change of the muon slope seen in generated events is still seen in
reconstructed events.

The mass composition assumed here is the JACEE one,
for all the Figs. 13 -- 15.
It should be noticed that for both energies only 10000 showers
were used to make that figures.

For the shower reconstruction there was assumed that the
detector are ''ideal''. This means that the detector response width
was set to 0. The fluctuations of the detector response are due
only to the extensive air shower developement physics.

\newpage

\centerline{\mbox{
\psfig{file=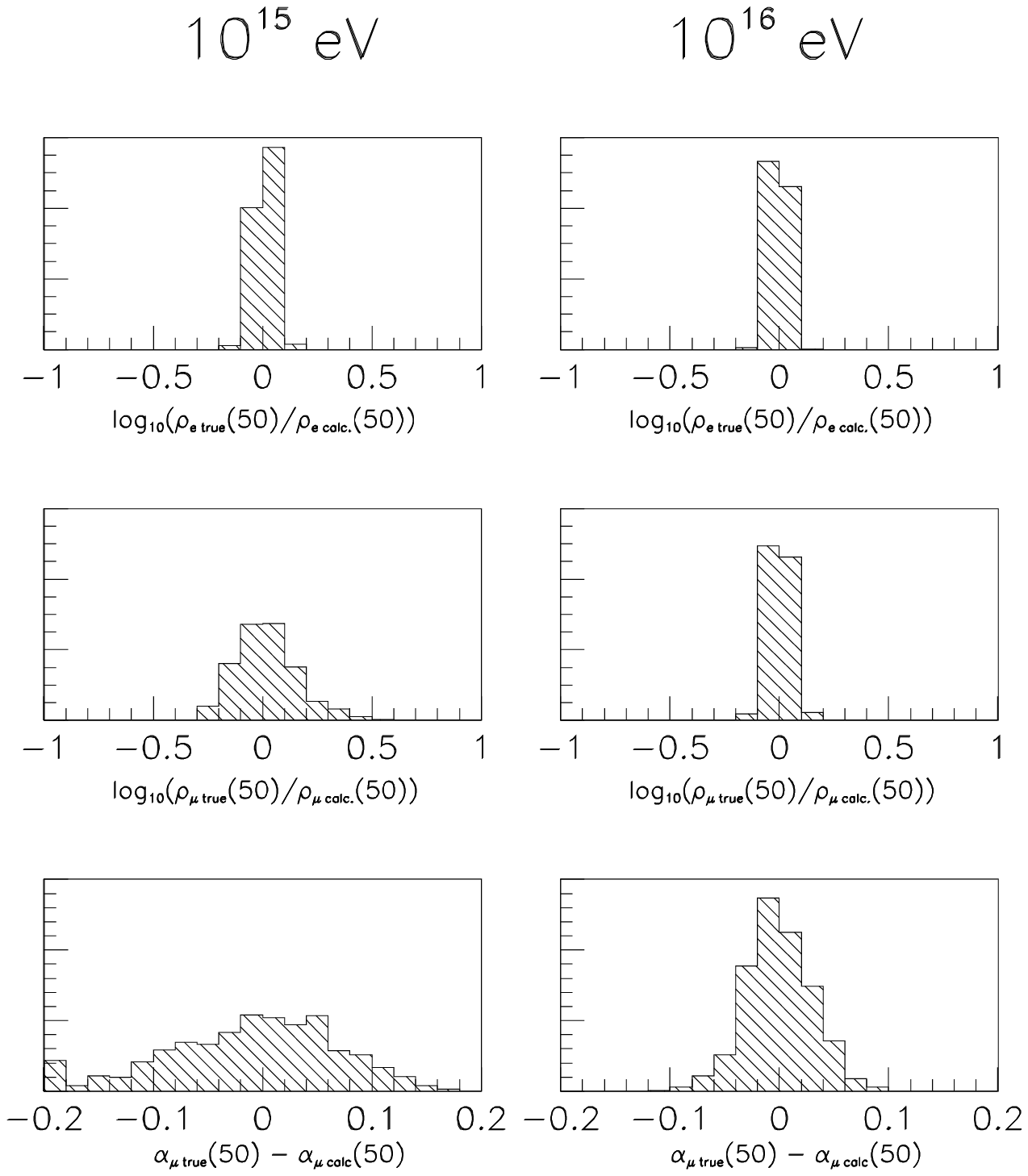,width=12cm,angle=0}
}}

Fig.14. The errors in the shower parameters determination.
The primary particle energy is
$10^{16}eV$
and
$10^{17}eV$ respectively.

\centerline{\mbox{
\psfig{file=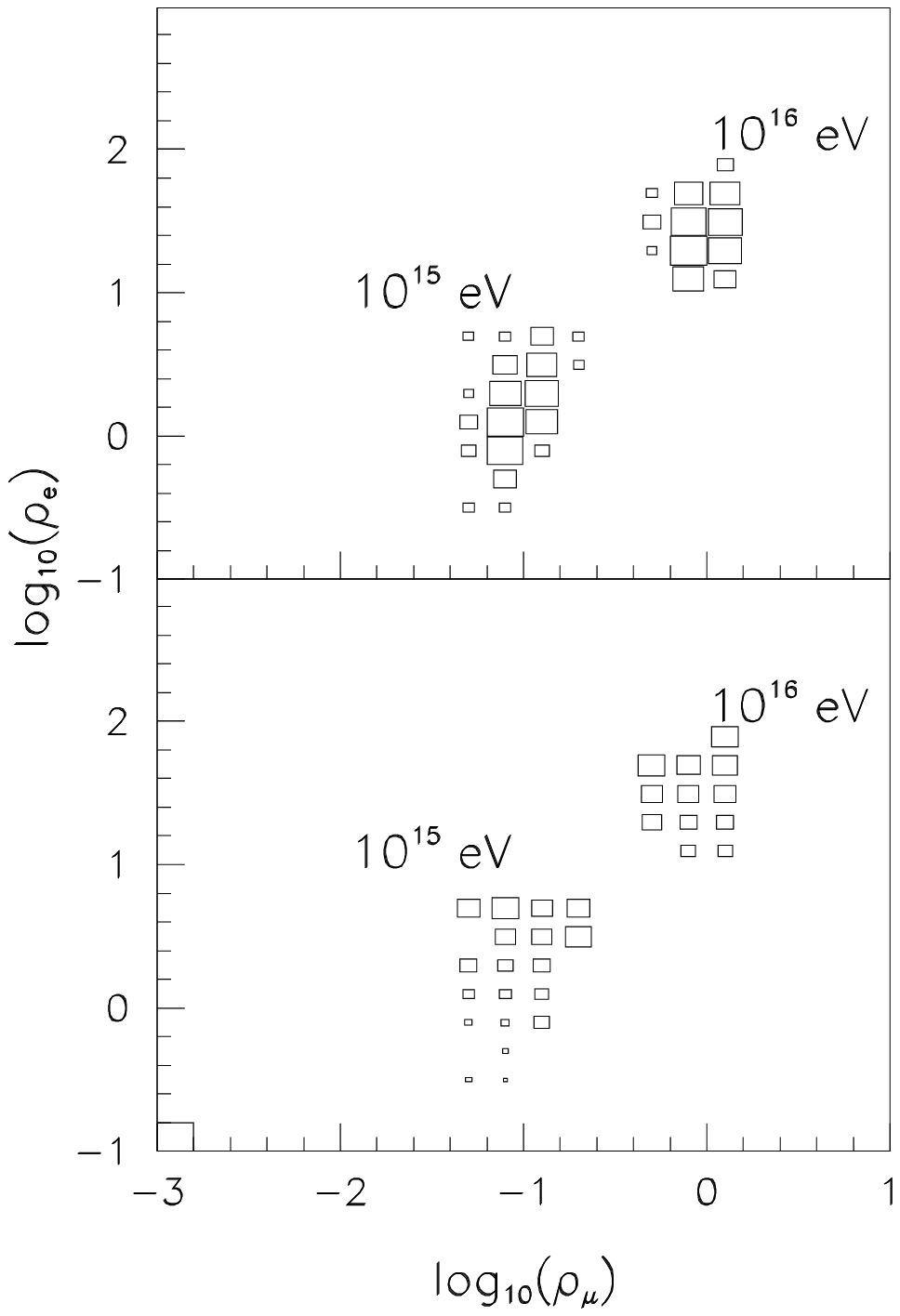,width=12cm,angle=0}
}}

Fig. 15.
The spectrum of $\left( \rho _{\mu}(10m), \rho_e (50m) \right)$.
and the slope of the muon distribution as it is given in
Fig. 8 for the showers used in Fig. 14.

\centerline{\mbox{
\psfig{file=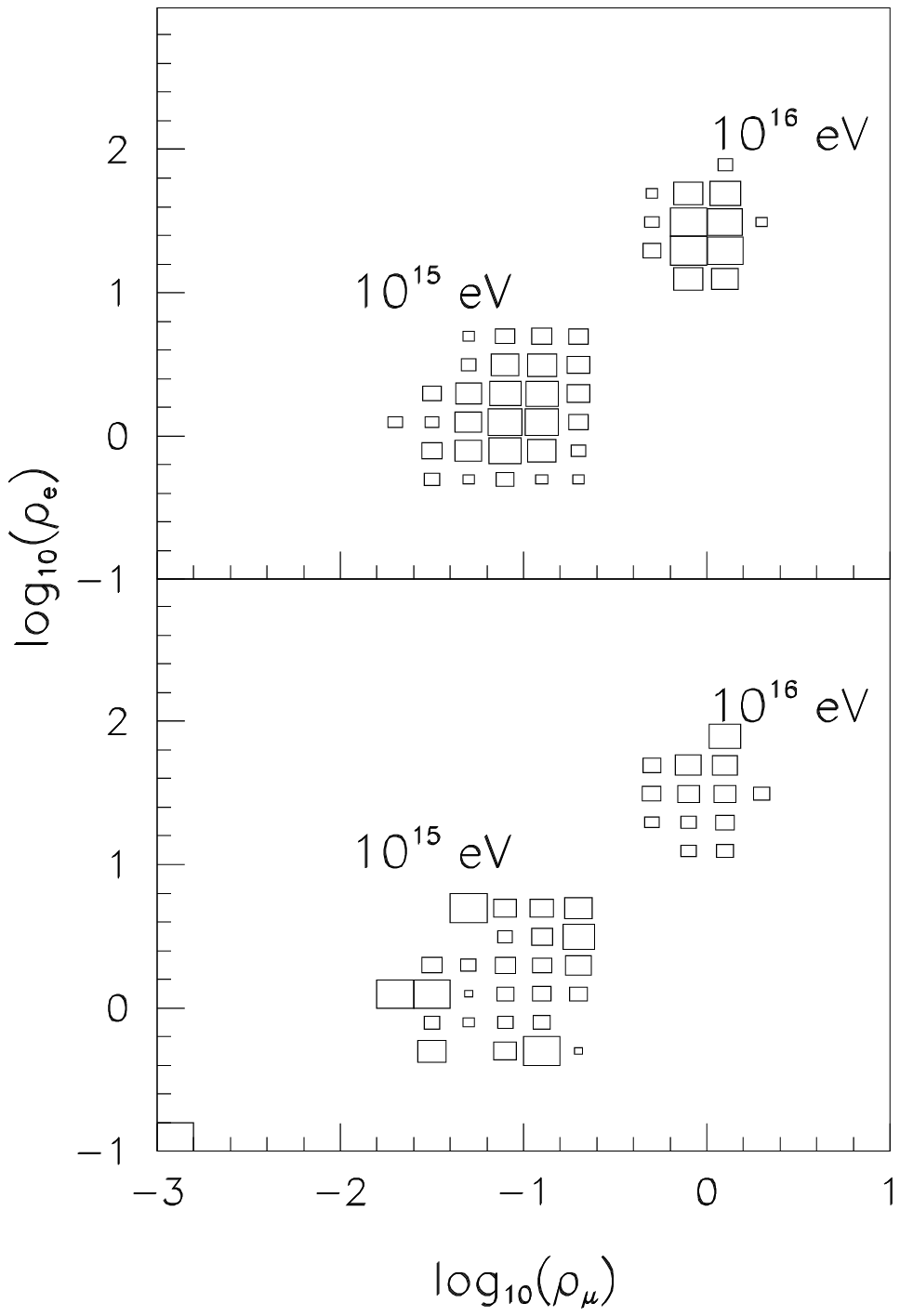,width=12cm,angle=0}
}}

Fig. 16.
The spectrum of reconstructed $\left( \rho _{\mu}(10m), \rho_e (50m) \right)$.
and reconstructed slope of the muon distribution as it is given in
Fig. 8 for the same showers as used in Fig. 15

\section {Acknowledgements}

I would like to thank Prof. H. Rebel for his useful suggestions and comments.


\begin{thebibliography}{9}

\bibitem {kascade} H. Rebel and KASCADE collaboration, ${\rm 7^{th}}$ ISVHCRI
Ann Arbor, AIP Conference Proc. ed. L. Jones, 575, (1992); G. Schatz,
Interdisc. Sci. Rev. {\bf 18}, 306 (1993).

\bibitem{sup} G. Schatz, T Thouw, K. Werner, J. Oerschlaeger and K. Beck,
J. Phys. G {\bf 20}, 1267 (1994).

\bibitem{jacee} K. Asakimori et al., Proc. ${\rm 24^th}$ ICRC Roma {\bf 2},
707 (1995).

\bibitem{corsika} J. N. Capdevielle et al., KfK Report 4998 (1992); J. Knapp
and D. Heck, {\it CORSIKA User's Manual}, Internal Report Forschungszentrum
Karlsruhe 1995 (unpublished).

\end{thebibliography}
\end{document}